# Band Structure Modulation in Ternary-Based $InAs/AlAs_{1-x}Sb_x/InAs_{1-x}Sb_x$ Superlattices

Arash Dehzangi, Jiakai Li

[1]Department of Electrical and Computer Engineering, University of Texas at Dallas, Richardson, Texas, USA.
[2]Applied Optoelectronics Inc, Sugar Land, Texas, 77478, USA

*Abstract*— **This study demonstrates an overall view of recent quantum mechanical based modeling effort using empirical tight-binding method to expand the ability of calculation of electronic band structure for strained-layer superlattices composed of ternary $InAs/InAs_{1-x}Sb_x$ and $InAs/InAs_{1-x}Sb_x/AlAs_{1-x}Sb_x$ materials. Using a modified sp3s* empirical tight-binding method and building the superlattice Hamiltonian, we demonstrate that combining the virtual crystal approximation with a bowing parameter for the s-on-site energy accurately predicts ternary superlattice behavior. We provide a comprehensive overview of our numerical simulations, which model diverse superlattice configurations ranging from long wavelength infrared range to wide-bandgap barrier architectures. Notably, the simulated energy levels strongly agree with experimental bandgap measurements of the ternary superlattices and their structural variants.**

## I. INTRODUCTION

While traditional state-of-the-art HgCdTe remains a benchmark, strained-layer superlattices (SLSs) have proven to be reliable alternatives for high-performance infrared photodetectors and focal plane arrays [1-7], for many different applications such as laser [8], remote sensing and space [9], high speed transistors [10-13].

The spatial separation of electrons and holes inherent to Type-II band alignment allows deterministic band structure engineering through localized thickness and compositional control of the alternating layers [14]. Transitioning from conventional but viable InAs/GaSb system [15-20], to InAs/ $InAs_{1-x}Sb_x$ (also known as Ga-free SLS) material systems has yielded critical benefits such as carrier lifetime and manufacturing scalability [21-24].

This design space is further expanded by the three-layer InAs/ $InAs_{1-x}Sb_x/AlAs_{1-x}Sb_x$ SLS, which introduces enhanced degrees of freedom with flexibility for tailoring mid-wavelength infrared (MWIR) and long-wavelength infrared (LWIR) material with barrier structures to improve the signal to noise ratio and control the dark current. InAs/InAs1-xSbx/AlAs1-xSbx SLS is more complex Ga-free type II superlattice structure that can include three different layer, allows for greater flexibility in engineering the energy bandgap through different combinations of layers and compositions for narrow band gap long wavelength infrared (LWIR) and mid wavelength infrared (MWIR) when barrier structures are needed [25, 26]. Structurally, however, this system lacks the favorable lattice-matching found in InAs/GaSb, since InAs and $InAs_{1-x}Sb_x$ are not closely lattice-matched. This could be even harder in strain balance, when it comes to $InAs/InAs_{1-x}Sb_x/AlAs_{1-x}Sb_x$ SLS. The resulting strain profile creates a great sensitivity to slight deviations in layer thickness or group-V composition (x). From a computational standpoint, this volatility introduces distinct challenges for electronic band structure modeling, particularly when simulating regimes with high antimony content (antimony-rich $x > 0.5$).

Some aspects of the $InAs/InAs_{1-x}Sb_x$ SLS have been modeled previously using k.p method [27-29]. Ab-initio calculation as a first principal approach, is capable of modeling SLS electronic band structure [30], yet it might suffer from approximation limitations, inaccuracy for modeling defects and segregations along with time-consuming procedure. The empirical tight-binding model (ETBM) is a semi-empirical, quantum mechanical method that utilizes a simplified atomistic approach relative to first-principles (ab initio) methods. A simple comparison between the ETBM and first-principles methodologies is summarized in Table 1.

The empirical nature of the ETBM provides an advantageous balance between computational efficiency and predictive accuracy compared to first-principles approaches. Despite its reduced computational complexity, the ETBM maintains high numerical accuracy when calculating the critical features of SLS electronic structures [31-34].

The framework begins by representing the material's wavefunctions as a linear combination of atomic orbitals (LCAO). To solve the Schrödinger equation, the Hamiltonian matrix elements—which define the electronic energy states—are treated as adjustable parameters. These parameters are systematically fitted to established experimental band structure data, such as characteristic bandgaps and effective masses. Once parameterized, the model enables computationally efficient simulations of larger, more complex architectures. Ultimately, the ETBM demonstrates strong capabilities for precisely modeling the electronic properties of SLS materials.

Despite its advantages, the ETBM presents specific limitations when applied to complex SLS systems. The precision of simulated energy bands depends heavily on the microscopic nature of the heterostructure interfaces. Modeling these regions with the ETBM can be challenging, as it requires a high volume of semi-empirical parameters to accurately capture interface mixing, monolayer thickness variations, strain fields, and localized defects. Some of these limitations along with spatial and charge ambiguities often emerge within the superlattice framework can ultimately degrade the accuracy of derived electronic properties, such as carrier velocities and effective masses.

Table 1. First-principles vs. semi-empirical methods [32, 35-37]

| Feature | First-Principles | ETBM |
|---|---|---|
| **Origin of parameters** | Parameters are determined from fundamental physical constants and equations, without fitting to experimental data. | Key parameters (e.g., hopping integrals, on-site energies) are treated as adjustable values and are fitted to reproduce known experimental data or the results of first-principles calculations. |
| **Computational cost** | Very computationally expensive, especially for large systems (e.g., over 1,000 atoms). | More computationally efficient, making it suitable for larger systems, including nanostructures and alloys. |
| **Accuracy and transferability** | Generally considered more accurate and reliable, as it can be used for a wide range of systems without modification. | Accuracy depends on the quality of the fit. The model may lack transferability and need to be refitted for new material systems or different structural configurations. |
| **Physical insights** | Provides a detailed quantum-mechanical description of electronic properties and interactions. | Offers a simplified, but often physically transparent picture of electronic structure and how it relates to atomic-orbital interactions. |

It has been successful implementation of ETBM approach for InAs/GaSb binary SLS, with reliable accuracy considering the growth conditions, cut-off wavelength, interface engineering and antimony segregation [38, 39]. As an example of the flexibility of ETBM we used similar approach explained in Ref [38, 40], calculated 3D map of cutoff wavelength for superlattice of $[(AsIn)_6\text{-}AsGa_{x1}In_{1-x1}\text{-}(SbGa)_{10}\text{-}SbGa_{x2}In_{1-x2}]_N$, as it is shown in Figure 1. The entire range for $x_1$ and $x_2$ (from 0 to 1) has been calculated for an energy variation of ~40 meV corresponding to cutoff wavelength ranges from 3.68 µm to 4.20 µm. By aligning this with calculated map of mismatch of suggested superlattice and the GaSb (001) substrate we can identify a zero-mismatch composition line.[38] This can lead to growth of high crystalline quality of superlattice design on GaSb substrate.

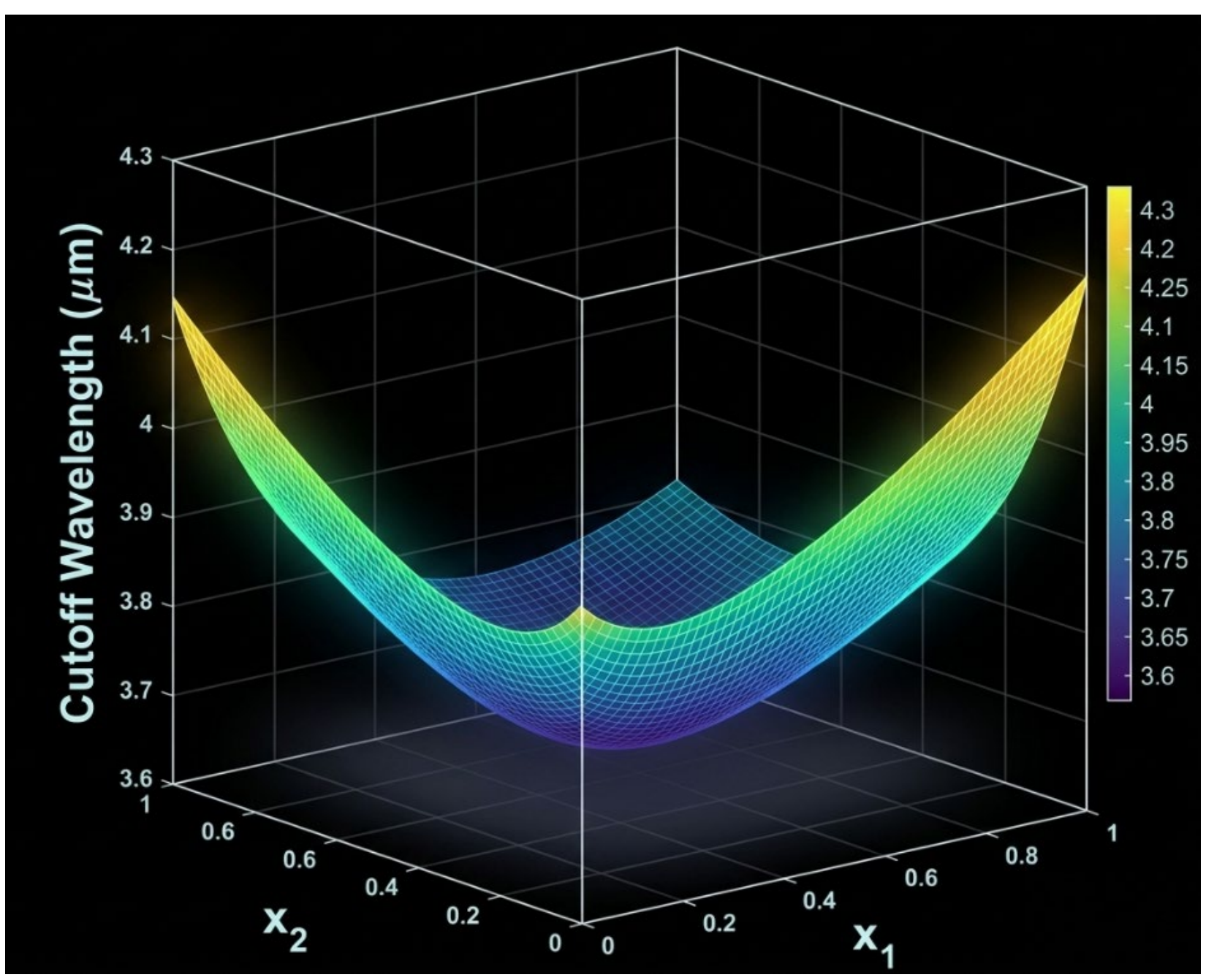

**Figure 1**. An example of ETBM modeling of superlattice design. The 3-D map of different cutoff wavelength for $[(AsIn)_6\text{-}AsGa_{x1}In_{1-x1}\text{-}(SbGa)_{10}\text{-}SbGa_{x2}In_{1-x2}]_N$, when $x_1$ and $x_2$ change from 0 to 1.

However, unlike InAs/GaSb binary superlattice, ETBM is not as straightforward for $InAs/InAs_{1-x}Sb_x$ or for $InAs/InAs_{1-x}Sb_x$ /$AlAs_{1-x}Sb_x$ ternary SLS. The random distribution of group-V elements in ternary compounds makes localizing individual As and Sb atoms difficult, complicating the fitting of nearest-neighbor orbital interaction energies [41-43]. Since these fitting parameters dictate the accuracy of the ETBM approach and are highly sensitive to material quality, a systematic parameterization strategy is required. Developing a universal rule is therefore necessary to reliably explain experimental results across superlattices fabricated under different growth conditions.

We have reported ETBM modeling approach for $InAs/InAs_{1-x}Sb_x$ material with detailed calculations and comparison with experimental data [44, 45]. In this work, broader view of using a modified sp3s* ETBM and virtual crystal approximation (VCA) approach on calculating electronic band structure of ternary-based Ga-free SLS with different design to cover from narrow band to wideband infrared regions. We also revisit the band structure modeling of narrow band InAs/ InAs1-xSbx SLS to give more details on lattice matching and strain compensation strategy with more systematic version of the evolution of our modeling. Experimental results for different SLS design were implemented to validate the approach.

## II. SUPERLATTICE HAMILTONIAN

Originally proposed to interpolate first-principles electronic structure calculations, the tight-binding method is rooted in Bloch's LCAO and then it has been refined over the years by various research groups [46]. A key advancement in constructing the Hamiltonian matrix for semiconductor SLS design was the orthonormalization of atomic orbitals [47], which successfully preserves structural symmetries. Building upon this framework, we utilize our previously proposed basis of sp3s* orbitals incorporating spin-orbit interactions [44]. The corresponding parameters for sp3s* tight-binding modeling have been thoroughly documented for both bulk III-V materials and III-V superlattices [48, 49]. Here we briefly describe our approach, more details about the approach can be found elsewhere [45]

Assuming that $\phi_n^\alpha$is a sp3s* Löwdin orbital of the atom in the crystal, Wave function is presented as follows, where n runs through all the atoms in a unit cell and α runs through all Löwdin orbitals (s, px, py, pz, and s*):

$$\psi_{\vec{k}}(\vec{r}) = \sum_{\overrightarrow{R_{cell}}} \sum_{n=1}^{N} \sum_{\alpha} C_n^\alpha \exp\left(i\vec{k}\cdot\left(\overrightarrow{R_{cell}}+\overrightarrow{\tau_n}\right)\right)\phi_n^\alpha\left(\vec{r}-\overrightarrow{R_{cell}}-\overrightarrow{\tau_n}\right) \quad (1)$$

where k is the wave vector, $\overrightarrow{R_{cell}}$ is a vector labeling for the position of one unit cell, $\tau_n$is the coordinate position vector of the atoms in the cell, and $C_n^\alpha$are constants. The energy dispersion relation of the crystal can be reduced to the following eigenvalue problem (H is the Hamiltonian):

$$H\psi_{\vec{k}}(\vec{r}) = E\psi_{\vec{k}}(\vec{r}) \quad (2)$$

We followed similar approach proposed in [38], assuming each layer of the SLS begins with an anion and ends with a cation. The Hamiltonian for superlattice is written:

$$H_{superlattice} = \begin{bmatrix} H_a & H_{ac} & 0 & 0 & \cdots & 0 & H_{ca}^+ \\ H_{ac}^+ & H_c & H_{ca} & 0 & \cdots & 0 & 0 \\ 0 & H_{ca}^+ & H_a & H_{ac} & 0 & \cdots & 0 \\ 0 & 0 & H_{ac}^+ & H_c & 0 & \cdots & 0 \\ \vdots & \vdots & 0 & 0 & \ddots & 0 & 0 \\ 0 & 0 & \vdots & \vdots & 0 & H_a & H_{ac} \\ H_{ca} & 0 & 0 & 0 & 0 & H_{ac}^+ & H_c \end{bmatrix} \quad (3)$$

where

$$H_v = \begin{bmatrix} h_{vv} & h_{v,so} \\ h_{v,so}^+ & h_{vv}^* \end{bmatrix}, \quad v = a, c \quad (4)$$

and

$$H_{ac} = \begin{bmatrix} h_{ac} & 0 \\ 0 & h_{ac} \end{bmatrix}, H_{ca} = \begin{bmatrix} h_{ca} & 0 \\ 0 & h_{ca} \end{bmatrix}, \quad (5)$$

All the parameters in Eq (3-5) were defined in Ref[44]. The interaction energies between nearest neighbor's orbitals were considered in SLS Hamiltonian, where $H_a$ and $H_c$ are inter-orbital orbital interaction and $H_{ca} = H_{ac}^*$is self-conjugated and is the intra-atomic orbital interaction matrix. While $H_a$and $H_c$are diagonal matrix because of the orthnormality of Löwdin's orbitals, the exact form of $H_{ac}$ has to be determined and can be greatly simplified by the symmetries of the crystal. Figure 2 shows schematic 3D representation of tetrahedral zinc blende lattice symmetry with cations and anions layout to illustrate interaction energies between nearest neighbor's orbitals considered in ETBM approach.

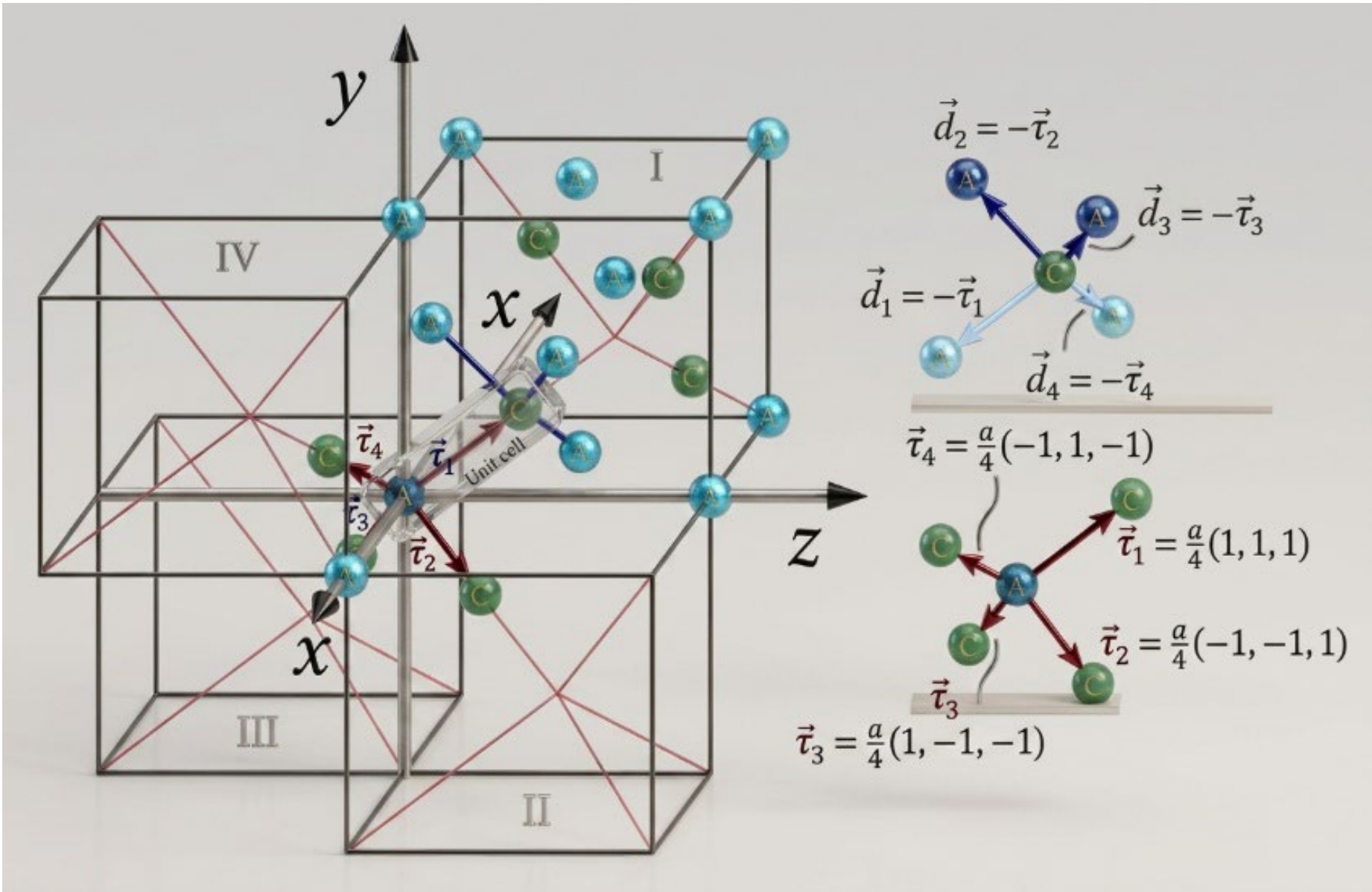

**Figure 2**. The 3D schematic of tetrahedral zinc blende lattice symmetry to illustrate interaction energies between nearest neighbor's orbitals

In the context of strain-balanced (strain-compensated) design, the balancing strain means that the tensile strain in one layer is compensated by a compressive strain in an adjacent layer, aiming for a net zero strain across the entire superlattice period. The goal is to manage the strain in a way to drastically reduce localized strain gradients that exist at the interfaces between layers, and keep it distributed throughout the superlattice period, rather than concentrated solely in the quantum wells [50]. In our approach the strain stems were extracted from the actual lattice mismatch between the substrate and the grown SLS material [51-58].
The empirical tight-binding parameters (TBPs) in our work are based on Vogl's and Klimeck's parameters [48, 59]. These TBP values represent the Hamiltonian matrix elements in the and are determined by fitting them to experimental data. In this work, the TBPs are mostly fitted for room temperature (300 K), and Harrison's d-2 scaling rule [60-62] was implemented to adjust them for modelling 77 K.

Table 2. The TBPs of different compound semiconductors at 77 K in eV.[44, 45]

| | InAs | InSb | AlAs | AlSb |
|---|---|---|---|---|
| $E_{sa}$ | -9.3562 | -9.3378 | -7.6201 | -4.5572 |
| $E_{sc}$ | -3.9611 | -3.3248 | -1.1786 | -4.1180 |
| $E_{pa}$ | +1.8201 | +0.39352 | +0.8905 | +0.01635 |
| $E_{pc}$ | +3.1842 | +2.0791 | +3.4939 | +4.8741 |
| $E_{s*a}$ | +7.0432 | +6.6378 | +7.3905 | +9.8429 |
| $E_{s*c}$ | +6.1232 | +5.3807 | +6.6339 | +7.4324 |
| $V_{sasc}$ | -6.5328 | -5.8320 | -6.7375 | -6.7066 |
| $V_{saxc}$ | +4.3563 | +4.1129 | +5.1668 | +4.6377 |
| $V_{xasc}$ | +7.0778 | +7.5769 | +6.3693 | +8.6279 |
| $V_{s*axc}$ | +2.9977 | +3.4448 | +4.5713 | +7.4657 |
| $V_{xas*c}$ | +5.3966 | +5.8873 | +7.3803 | +6.3653 |
| $V_{xaxc}$ | +2.5466 | +1.2596 | +1.8987 | +1.1192 |
| $V_{xayc}$ | +5.4646 | +4.0026 | +3.9025 | +4.9535 |
| $\Delta_a$ | +0.420 | +0.973 | +0.350 | +0.704 |
| $\Delta_c$ | +0.393 | +0.393 | +0.024 | +0.0306 |

At lower temperature higher interaction energy is expected between the covalently bonded orbitals. For that purpose, the TBPs have been fitted by multiplication of the inter-atomic orbital interaction energies [39], to reflect the impact of temperature on the mean distance between a given anion/cation couple. Varshni's empirical law [63] was considered in our approach to implement the impact of temperature on the simulated band gap around the Γ-point. The TBP values at 77 K for different compounds are presented in Table 2. The interaction energies for the s, p and s* orbitals, which can be used in the SLS Hamiltonian, are presented as Vsasc, Vsaxc, Vxasc, Vs*axc, Vxas*c, Vxaxc, Vxayc. In Table 1, Esa, Esc, Epa, Epc, Es*a, Es*c are relative energy values fitted to set the top of the valence band at a reference energy origin. These values are necessary to find accurate band alignment in the modeling, especially for valence band offset (VBO). Different approaches have been suggested for VBO calculation such as ab-initio [64] or direct extraction from the difference in the electron affinities of the two materials in the hetero-junction [65].
Some theoretical calculations for band offset have been suggested on bulk band structure on an absolute energy scale. In this work, we follow the approach proposed by Van de Walle and Martin [66] as presented in Table 3 for different compound materials. declared.

Table 3. The VBO of relevant compound materials used in the superlattice Hamiltonian [45, 67]

| | InAs | InSb | AlAs | AlSb |
|---|---|---|---|---|
| VBO (eV) | -6.543 | -5.82 | -7.40 | -6.44 |

## III. TIGHT BINDING PARAMETERS FOR TERNARY COMPONENTS

When it comes to ternary alloys with more complex design such as InAs/InAs1-xSbx and InAs/InAs1-xSbx/AlAs1-xSbx SLS, more precise modeling is needed for the band structure. Aside from complexity, major challenges are to identify the thickness of each layer and molar fractions (x) of the group V for all the constituent layers in the ternary structure. At the first step, we started to use a direct method to generate the TBPs of bulk ternary alloys before including these alloys in the superlattice Hamiltonian. VCA was used to model the atomic composition of the ternary alloy, by replacing the two substitutable group V atoms in a ternary bulk alloy by a virtual atom with shared weighted average properties of the two initial group V atoms. In terms of TBPs, we define them for ternary alloys as the weighted average of the two binary compound semiconductors, such as an extension of Vegard law [44].
There is an issue with Vegard's law [58] for the bandgap approximation for ternary alloys; due to bowing effect (parabolic or more complex shape) in the values of bandgap as a function of the composition. This nonlinearity of the bandgap of ternary alloys generate complexity in modeling based on tight-binding method. While the exact reason for this non-monotonic behavior is still debatable, some reports connect this to disorder effects;[68, 69] or to other properties of the band structure of the initial components such as VBO or the lattice mismatch [70]. The non-linear behavior of the simulated bandgap was accurately addressed in our previous calculation for $InAs_{1-x}Sb_x$ [44]. It is worth adding that the unwanted bowing can be explained by looking closely at the theoretical expression of the bandgap around the Γ-point using the tight-binding method:

$$E_c = \left[\frac{E_{sa}+E_{sc}}{2}\right] + \sqrt{\left[\frac{E_{sa}-E_{sc}}{2}\right]^2 + V_{sasc}^2}, \qquad (6a)$$

$$+\sqrt{[\frac{E_{pa}-E_{pc}+\Delta_a-\Delta_c}{2}]^2 + V_{xaxc}^2} \qquad (6b)$$

where the terms $V_{sasc}^2$ and $V_{xaxc}^2$ inside the square roots are responsible of the non-linearity of $E_c$ and $E^v$ as a function of the group V molar fraction (x) when the parameters $E_{sa}$, $E_{sc}$, $E_{pa}$, $E_{pc}$ are linearly interpolated for ternary alloys explained elsewhere [44].
Following the approach explained in Ref[44], the bowing parameter $b_v$ in the VCA can be written as [71]:

$$b_v = k\frac{|E_{sv}^{AC}-E_{sv}^{BC}|^{\lambda}}{|V_{sasc}^{AC}-V_{sasc}^{BC}|} \qquad (7)$$

This introduces k and λ as unknown parameters. If the two binary compound semiconductors which are the basis of a ternary alloy share the same s-on-site energy, the bowing of the energy gap should be nearly equal to zero, which agrees with experimental results. The bowing parameter $b_v$ in the VCA with linear, and second order bowing for the s-on-site energies have been reported [72] and fitted to the commonly agreed values of the bandgap energy at Γ-point of ternary compounds. We considered λ=1 in our calculations, the resulting fitted values for $InAs_{1-x}Sb_x$, $GaAs_{1-x}Sb_x$, $AlAs_{1-x}Sb_x$, $Ga_xIn_{1-x}As$, $Ga_xIn_{1-x}Sb$, $Al_xIn_{1-x}As$, and $Al_xIn_{1-x}Sb$ have been calculated [44] .
Using these data, Figure 3a,b represent the predicted bandgap energies at the Γ-point for $InAs_{1-x}Sb_x$ and $AlAs_{1-x}Sb_x$ bulk ternary alloys. The blue lines correspond to the reference bandgap values, while the purple and green circles representing simulated values using VCA with and without second order bowing, respectively. By using both VCA and second order bowing parameter for the s-on-site $E_{sv}$ energy an agreement with experimental data was achieved (purple circles). An interesting observation is to see how the linear weighted average of the tight-binding parameters can partially include a bowing of the band gap more for the $AlAs_{1-x}Sb_x$ and much less for $InAs_{1-x}Sb_x$ ternary alloy. This validates the choice of implementing bowing the s-on-site $E_s$ energy for tight-binding method modeling of III-V ternary compound.

## IV. Impact of Sb Segregation Impact On Band Structure Modeling

To give a better view of modeling InAs/InAsSb and InAs/InAsSb/AlAsSb SLS grown by MBE in a realistic condition, the impact of segregation of superlattice many layers need to be taken into calculations for theoretical modeling. Segregation is one of the side effects of MBE growth of SLS heterojunctions when sometimes element mixing with each other which in turn can change the electronic properties of the grown superlattice [73-78]. This is even more pronounced for ternary SLS system like InAs/InAsSb/AlAsSb. During growth of ternary SLS by MBE, the samples surface getting sprayed by molecular beam, which is composed of thermally activated particles. Some of these particles following statistical processes remain inside the growth chamber, make them prone to be get incorporated in subsequent growth periods and introduce the segregation to the SLS material. Given the nature of segregation as thermal excitation of atoms, two-state Boltzmann distribution was implemented where a given atom is whether a free particle or tightly bonded on the sample.

The model was developed using the Boltzmann factor to qualitatively determine which elements are more prone to segregate[44]. Since our ETBM approach is based on actual growth condition, we assume the group V is major contributor to segregation, since the MBE growth is group V rich and group III limited. Between Sb and As for group V elements in our SLS growth, since the bond strength of Sb with group III in our case is weaker than that of As, we also assumed that Sb in the only segregated atoms in SLS structure. This is not far from reality since many other experimental and modeling data supported this assumption [78-80]. More details about the implementing Sb segregation in our ETBM approach were reported previously[44]. It is important to mention, it is expected the impact of segregation approximation would be less pronounced compared to the accuracy of choices were given for TBPs or VBO.

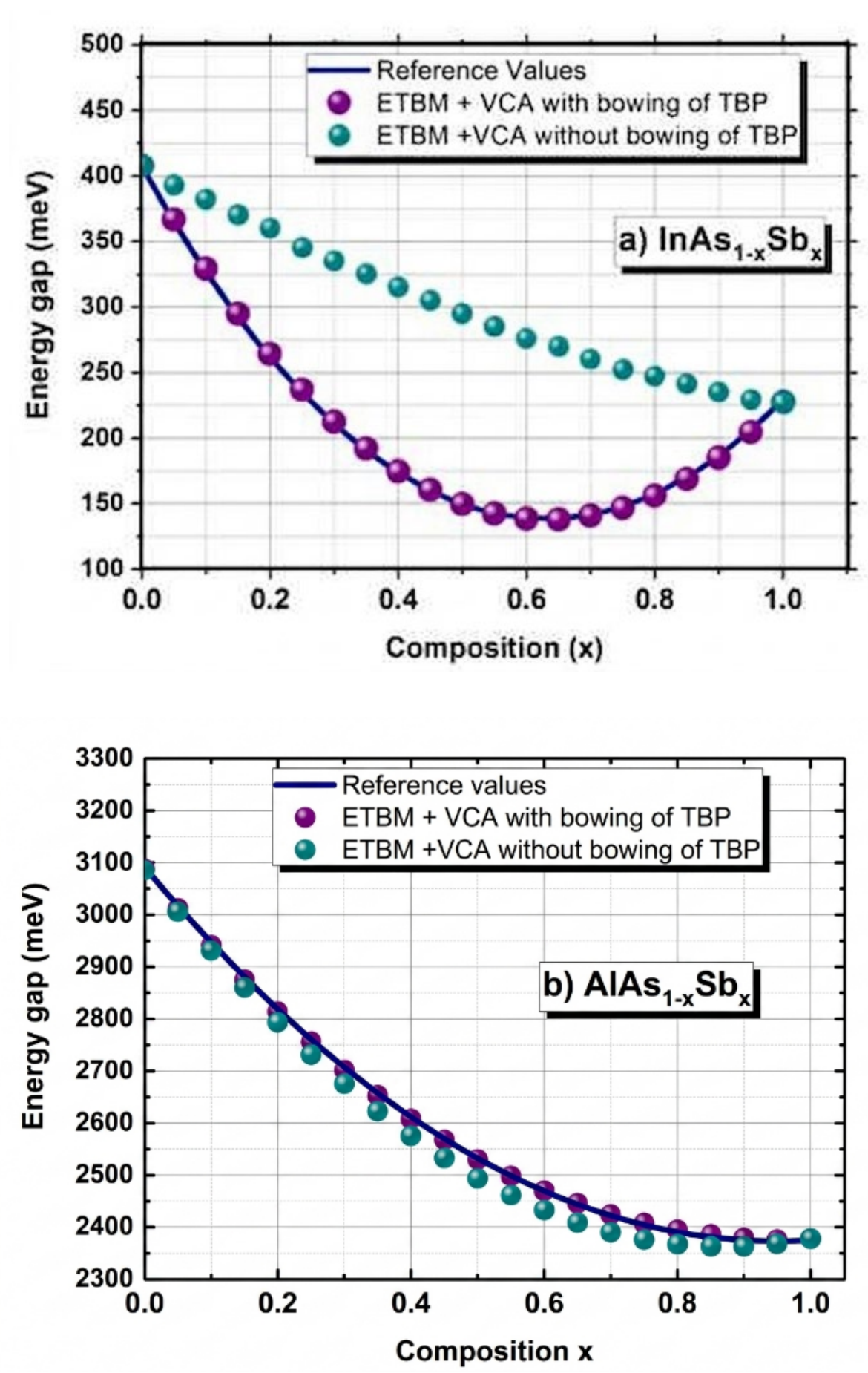

Figure 3. Simulated bandgap of (a) $InAs_{1-x}Sb_x$ and (b) $AlAs_{1-x}Sb_x$ III-V bulk ternary alloys. The red lines are the reference bandgap values. The green and orange circles correspond to simulated values using VCA with and without second order bowing respectively

is important to mention, it is expected the impact of segregation approximation would be less pronounced compared to the accuracy of choices were given for TBPs or VBO.

## V. Results for InAs/InAsSb/AlAsSb superlattice

We previously performed bandgap modeling of LWIR InAs/InAs1-xSbx. For the InAs/InAs1-xSbx type-II superlattices (Figure 4a), the electrons and holes are spatially separated in InAs and InAs1-xSbx quantum wells and the effective bandgap of the superlattice can be engineered by changing the layer thicknesses and antimony molar fraction of the InAs1-xSbx layer. It has been demonstrated that the bandgap of this type of superlattices can be widely tuned from a few tens to a few hundreds of meV.
Figure 4b presents the simulation results for [(InAs)35 – (InAs0.57Sb0.43)12] superlattice design which is in a good agreement with previous experimental work [81] for identical design. Other comparison with experimental results confirming the same outcome. The modelling calculation shows 83 meV for the bandgap, which is corresponding to 14.93 µm at 77 K, which is very close to 50% cut-off frequency reported in the experimental result.

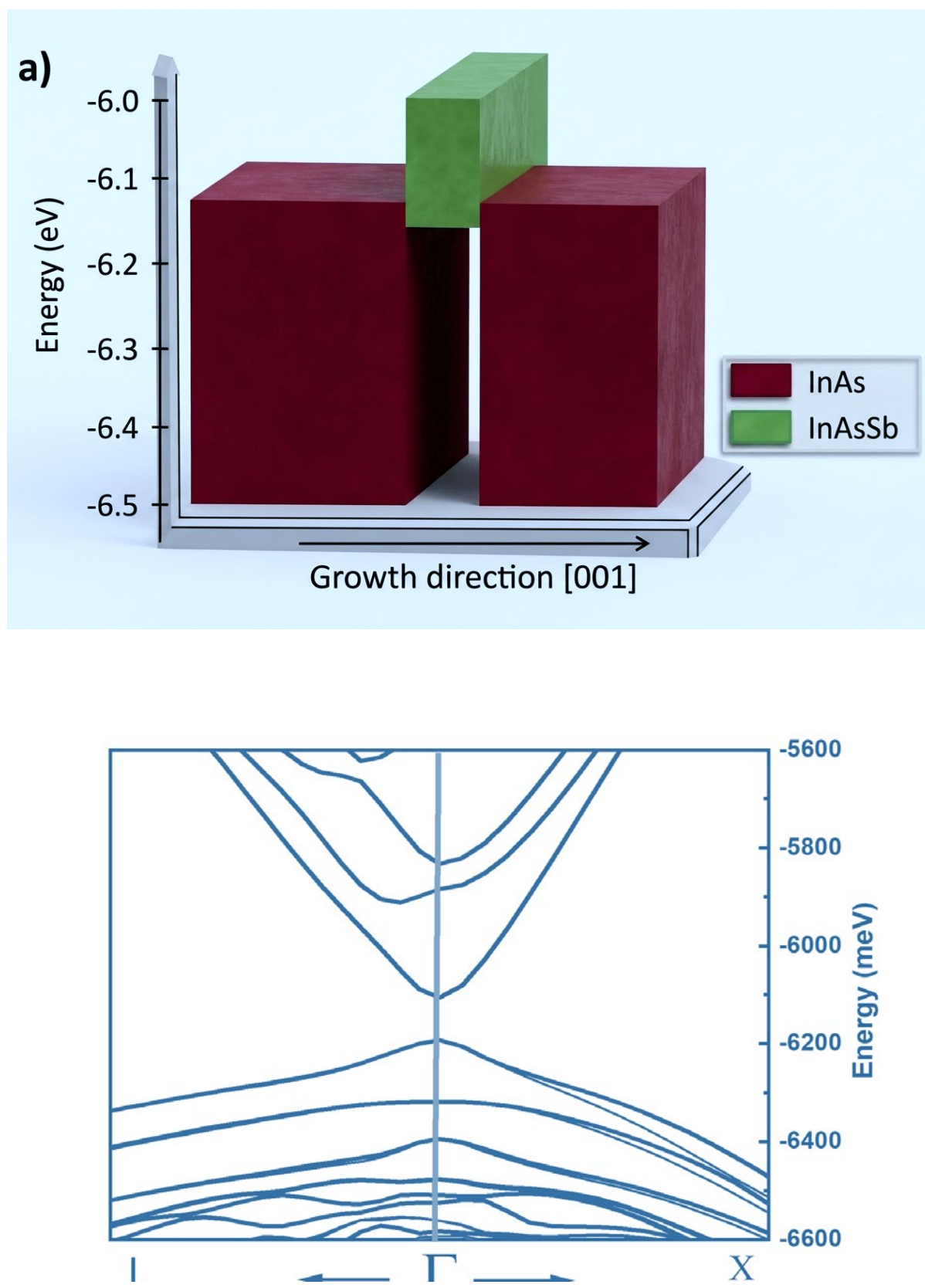

Figure 4. (a) [$(InAs)_{35}$ – $(InAs_{0.57}Sb_{0.43})_{12}$] superlattice design and (b) simulated band structure around the Γ-point of the superlattice using a *$sp^3s$** tight-binding model. The colored regions represent the energy gap of each layer in the structure.

## VI. Results for InAs/InAsSb/AlAsSb superlattice

For structures based on InAs/ $InAs_{1-x}Sb_x$/$AlAs_{1-x}Sb_x$, we aimed for wider band gap material using a barrier structure. As we gained good agreement on the choice of TBPs with experimental results (Figure 4). We also implemented the model to predict the band structure of several different superlattice designs based on ternary InAs/ $InAs_{1-x}Sb_x$/$AlAs_{1-x}Sb_x$. Figure 5 shows a comparison of the predicted bandgap energy using our ETBM model with actual experimental values for SLS design with short wavelength infrared (SWIR) and MWIR range at 77 K. All the data for experimental data was taken from MBE grown samples across the different range of the wavelength. We assumed that the segregation pattern is the same for all samples.

In InAs/$InAs_{1-x}Sb_x$ type-II superlattices design the electrons and holes are spatially separated in InAs and $InAs_{1-x}Sb_x$ quantum wells. This gives a flexibility to design engineering the effective bandgap of the superlattice by changing the Sb molar fraction, or thickness, type and sequence of each layer [51]. For that purpose, the idea was to exploit the high band gap at Γ-point of AlAs, Eg (AlAs)=3.09 eV and AlSb, Eg (AlSb)=2.37 eV by inserting a layer of high band gap ternary AlAs1-xSbx alloy between each quantum well InAs superlattice resulting in a barrier structure of InAs/$AlAs_{1-x}Sb_x$/InAs/$InAs_{1-x}Sb_x$ structure. Figure 6a shows schematic design of InAs/$AlAs_{1-x}Sb_x$/InAs/$InAs_{1-x}Sb_x$ SLS where an $AlAs_{1-x}Sb_x$ wider band gab material is inserted in the middle of the InAs quantum well as a barrier to elevate the effective conduction band energy. The $AlAs_{1-x}Sb_x$ barrier forms a double quantum well for electrons in the InAs layer that can increase the effective bandgap of the superlattice. This superlattice design inherits the properties of the InAs/$InAs_{1-x}Sb_x$ type-II superlattice as shown in Figure 6b. In particular, one can increase the bandgap by reducing the thickness of InAs quantum well.

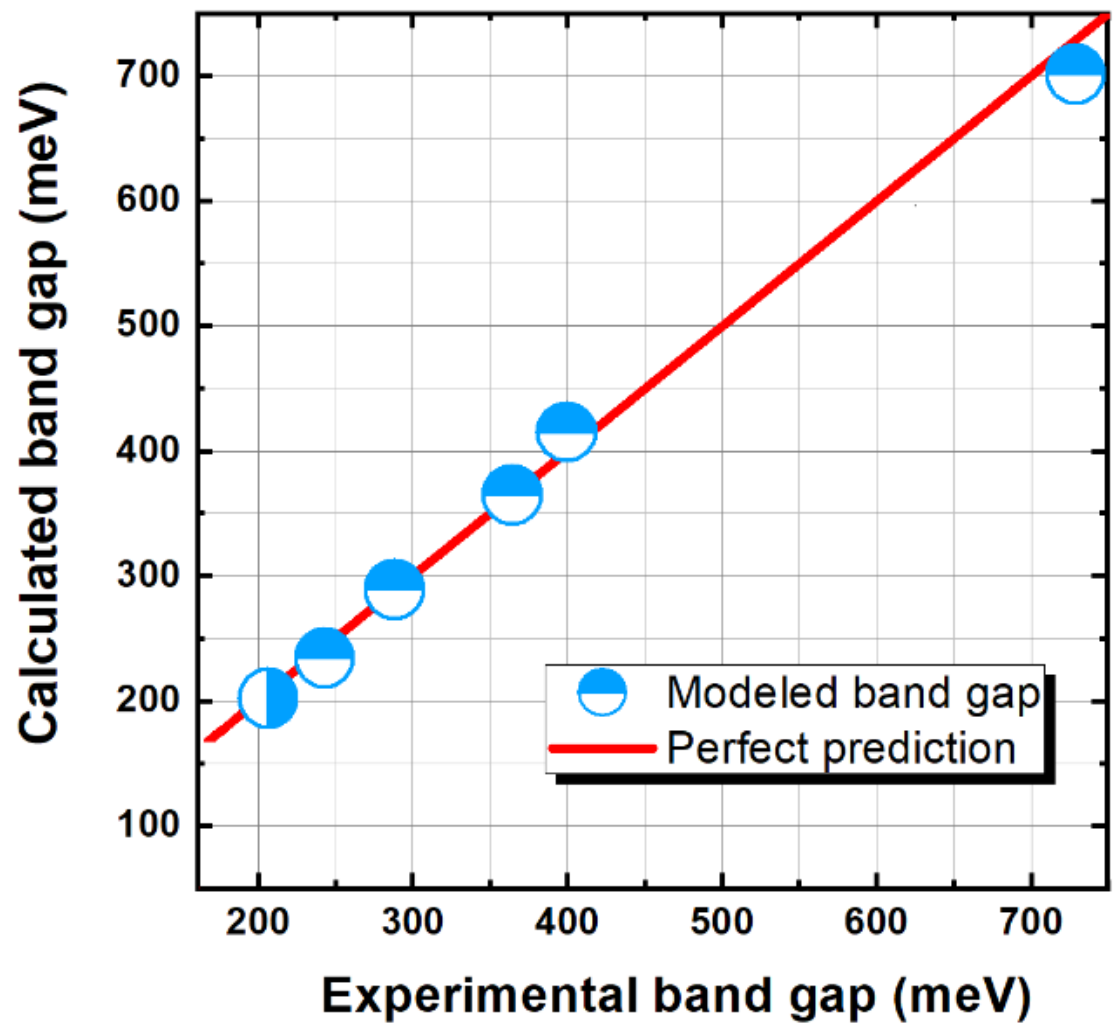

**Figure 5**. Comparison between the simulated bandgap of different superlattice designs of ternary alloys and measured bandgap energy of corresponding experimental superlattices grown with MBE.

The bandgap of this superlattice design increases by using a thicker $AlAs_{1-x}Sb_x$ barrier [82]. This specific design representing $[(InAs)_4–(AlAs_{0.52}Sb_{0.48})_2–(InAs)_4– (InAs_{0.52}Sb_{0.48})_7]$ superlattice design aligned with previous experimental work [82, 83]. The modelling calculation shows ~260 meV for the bandgap, which corresponds to 5 µm at 77 K, which is close to cut-off frequency reported in the experimental result.

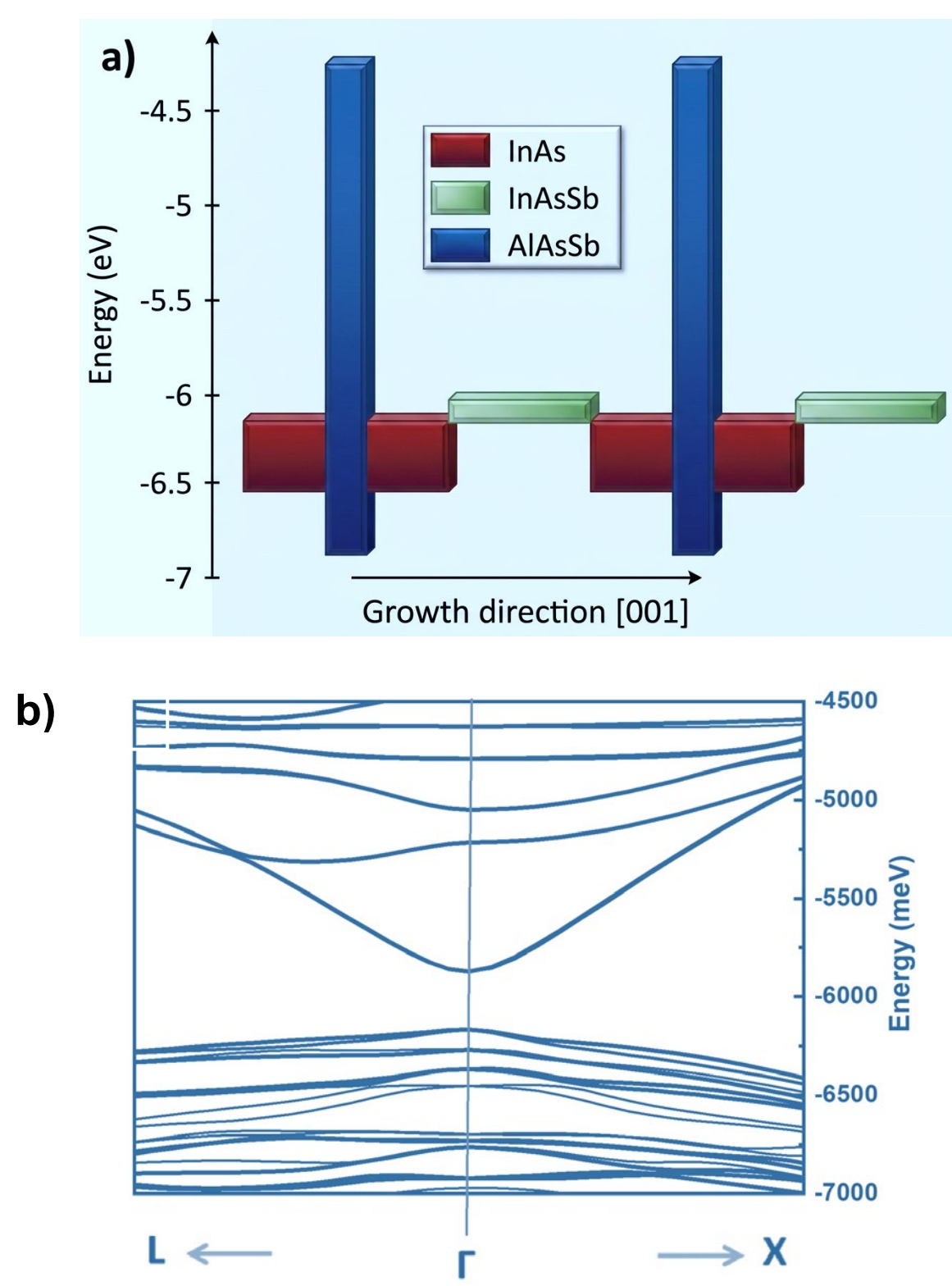

Figure 6 (a) $[(InAs)_4 – (AlAs_{0.52}Sb_{0.48})_2 – (InAs)_4 – (InAs_{0.52}Sb_{0.48})_7]$ SLS design, (b) simulated band structure around the Γ-point of the SLS using a $sp^3s^*$ tight-binding model. The colored regions represent the energy gap of each layer in the structure.

## VI. SHORT WAVE INFRARED DOUBLE BARRIER STRUCTURE

We pushed the design to cover the short wavelength infrared (SWIR) region. The first step is to incorporate binary or ternary high band gap layer to embed into the SLS design. Even if AlAs has a greater band gap than AlSb, the lattice mismatch between AlAs and the standard substrate GaSb (appropriate substrate for the growth) is too high (7%) to grow thick layers of AlAs on GaSb substrates. Instead, we implemented the ternary alloy $AlAs_{1-x}Sb_x$, since AlSb is lattice matched with GaSb. This way, both the high band gap of AlAs and the growability of AlSb are exploited. Our calculation for material growth, revealed ternary $AlAs_{1-x}Sb_x$ has lower mismatch and is more appropriate for generating a barrier inside the SLS structure.

In this structure we have $InAs/AlAs_{1-x}Sb_x/InAs/InAs_{1-x}Sb_x/AlAs_{1-x}Sb_x/InAs_{1-x}Sb_x$ SLS where $AlAs_{1-x}Sb_x$ barriers are inserted at the middle of the InAs and $InAs_{1-x}Sb_x$ quantum wells (Figure 6a). The $AlAs_{1-x}Sb_x$ barrier in both InAs/InAsSb SLS forms a double barrier structure in SLS design that can increase the effective bandgap of the superlattice.

Table IV shows the results for our calculations on multiple designs based on the $InAs/AlAs_{1-x}Sb_x/InAs/InAs_{1-x}Sb_x/AlAs_{1-x}Sb_x/InAs_{1-x}Sb_x$ SLS structure. The mismatch with GaSb substrate of each design is given and is calculated as the ratio of the average lattice constant ($a_{AV}$) of the superlattice over the lattice constant of GaSb ($a_{GaSb}$):

$$mismatch = \frac{a_{AV}}{a_{GaSb}} = \frac{1}{a_{GaSb}} \frac{\sum_{n=1}^{N} t_i a_i}{\sum_{n=1}^{N} t_i} \quad (12)$$

where $t_i$ is the thickness of layer $i$, $N = 6$ is the total number of layers in the $InAs/AlAs_{1-x}Sb_x/InAs/InAs_{1-x}Sb_x/AlAs_{1-x}Sb_x/InAs_{1-x}Sb_x$ SLS structure, and $a_i$ is the lattice constant of layer $i$, using Vegard's law for ternary alloys. The calculated band gap is also given for each structure.

**Table IV** Different designs of SLS structure with $AlAs_{1-x}Sb_x$ barriers for the band gap in $meV$. The designs are given in number of monolayers. Mismatches are given in $ppm$ (parts per million)

| Design (monolayers) | $Eg$ with $AlAs_{1-x}Sb_x$ ($meV$) | Mismatch ($ppm$) |
|---|---|---|
| 3/2/3/3/2/3 | 531.4 | -1462 |
| 2/2/2/3/2/3 | 582.3 | -2439 |
| 3/2/3/3/3/3 | 582.4 | +562.0 |
| 1/2/1/3/2/3 | 637.6 | -3736 |
| 3/3/3/3/3/3 | 635.6 | +2375 |
| 2/2/2/3/3/3 | 639.8 | -87.43 |
| 3/2/3/3/3/3 | 696.2 | +1988 |
| 1/2/1/3/3/3 | 701.1 | -934.3 |
| 1/3/1/3/3/3 | 758.5 | +1493 |

In Table 4, for all the designs, the same thickness of $InAs_{1-x}Sb_x$ quantum well was intentionally taken. Having the same thickness of $InAs_{1-x}Sb_x$ layer, allows us to keep the same valence band alignment for all the designs. As expected, the band gap of the designs increases as the InAs quantum well layer is reduced [84, 85], while the band gap rises with the thickness of each high band gap $AlAs_{1-x}Sb_x$ barrier.

More importantly, comparing the lattice mismatch with substrate for different designs led us to select the appropriate structure with the highest bandgap for SWIR region. Without considering the elasticity properties of crystals involved for critical thickness, it has been shown that an acceptable lattice mismatch for SLS structure on GaSb substrate is less than 1000 ppm [51, 86]. Given to the highest bandgap and mismatch consideration (< 1000 ppm) the final design chosen for the growth was the 1/2/1/3/3/3 design. The result is illustrated in Figure 8.

This superlattices design helps to separate the effective conduction and valence bands of the superlattice from each other as much as possible, and allows achieving high bandgap energies ($E_g$=701.1 meV, Figure 7b and Table 4) corresponded to cut-off frequency of 1.9 µm and close to result reported in the experimental results [87, 88].

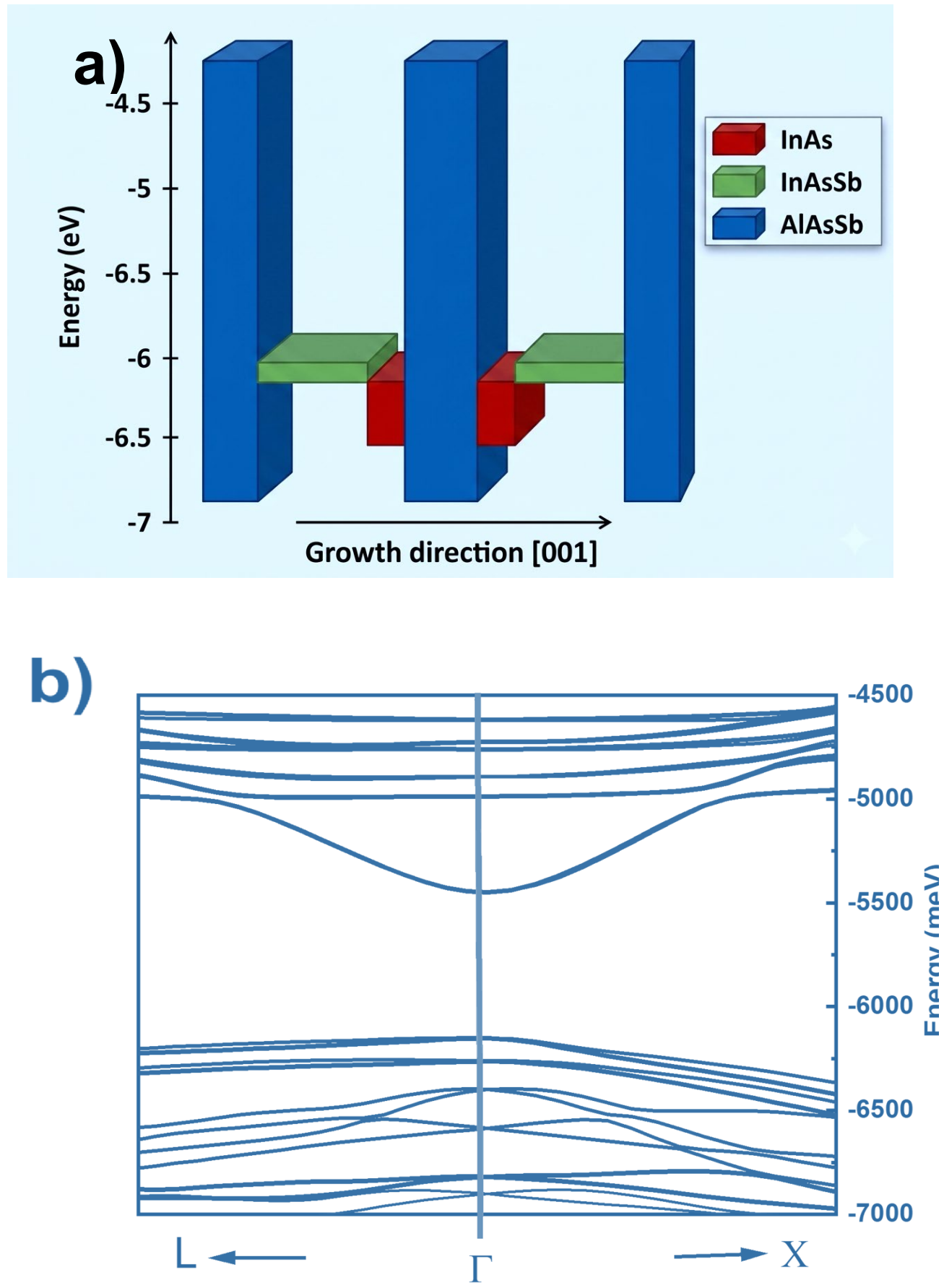

**Figure 7.** (a) [$(InAs)_1$–$(AlAs_{0.52}Sb_{0.48})_2$–$(InAs)_1$ – $(InAs_{0.52}Sb_{0.48})_3$ –$(AlAs_{0.52}Sb_{0.48})_3$ – $(InAs_{0.52}Sb_{0.48})_3$] superlattice design and (b) simulated band structure around the Γ-point of the superlattice using a $sp^3s^*$ tight-binding model. The colored regions represent the energy gap of each layer in the structure.

## VI. Conclusion

In conclusion, this study demonstrates that the modified ETBM approach reliably models the bandgap and cutoff wavelength of InAs/InAs1-xSbx and InAs/InAs1-xSbx/AlAs1-xSbx ternary SLS material across a broad range of infrared spectrum. By integrating the virtual crystal approximation to determine precise fitting parameters and accounting for antimony segregation during superlattice growth, the model achieves good accuracy. The ETBM simulations show strong agreement with experimental data across various superlattice designs, validating the appropriateness of this framework for designing more complex architectures in future research.

## Acknowledgment

The authors would like to thank Prof. Manijeh Razeghi the director of center for quantum devices (CQD) at Northwestern University. Authors also thank Dr. Abbas Haddadi, Dr. Donghai Wu and Xun Victor Suo. The MBE grown was performed at CQD.

## References

[1] G. A. Sai-Halasz, R. Tsu, and L. Esaki, "A new semiconductor superlattice," Applied Physics Letters, vol. 30, no. 12, pp. 651–653, 1977, doi: 10.1063/1.89273.
[2] A. M. Hoang, A. Dehzangi, S. Adhikary, and M. Razeghi, "High performance bias-selectable three-color Short-wave/Mid-wave/Long-wave Infrared Photodetectors based on Type-II InAs/GaSb/AlSb superlattices," Scientific Reports, Article vol. 6, p. 24144, 04/07/online 2016, doi: 10.1038/srep24144.
[3] M. Razeghi, A. Dehzangi, and J. Li, "Multi-band SWIR-MWIR-LWIR Type-II superlattice based infrared photodetector," Results in Optics, vol. 2, p. 100054, 2021/01/01/ 2021, doi: https://doi.org/10.1016/j.rio.2021.100054.

[4] A. Dehzangi, A. Haddadi, R. Chevallier, Y. Zhang, and M. Razeghi, "Fabrication of 12 µm pixel-pitch 1280 × 1024 extended short wavelength infrared focal plane array using heterojunction type-II superlattice-based photodetectors," Semiconductor Science and Technology, vol. 34, no. 3, p. 03LT01, 2019/02/04 2019, doi: 10.1088/1361-6641/aaf770.
[5] A. Dehzangi "Gain based device for long-wavelength infrared detection." In *Image Sensing Technologies: Materials, Devices, Systems, and Applications XIII*, vol. 14023, p. 9. SPIE, 2026
[6] M. Bouschet et al., "Transport properties in InAs/InAsSb type-II superlattices: From MWIR to VLWIR," Applied Physics Letters, vol. 126, no. 24, 2025, doi: 10.1063/5.0265432.
[7] A. Dehzangi, J. Li, L. Gautam, and M. Razeghi, "Avalanche Photodetector Based on InAs/InSb Superlattice," Quantum Re-ports, vol. 2, no. 4, pp. 591–599, 2020.
[8] S. Zhang et al., "Effect of the V/III ratio on the quality of strain-balanced GaInAs/AlInAs superlattices in quantum cascade lasers," Optical Materials, vol. 162, p. 116882, 2025/05/01/ 2025, doi: https://doi.org/10.1016/j.optmat.2025.116882.
[9] D. Z. Ting et al., "Type-II Superlattice Mid-Wavelength Infrared Focal Plane Arrays for CubeSat Hyperspectral Imaging," IEEE Photonics Technology Letters, vol. 34, no. 6, pp. 329–332, 2022, doi: 10.1109/LPT.2022.3156048.
[10] A. Dehzangi et al., "High speed antimony-based superlattice photodetectors transferred on sapphire," Applied Physics Ex-press, vol. 12, no. 11, p. 116502, 2019/10/17 2019, doi: 10.7567/1882-0786/ab4a8e.
[11] J.-H. Tsai, W.-S. Lour, D.-F. Guo, W.-C. Liu, Y.-Z. Wu, and Y.-F. Dai, "InP/GaAsSb type-II DHBTs with GaAsSb/lnGaAs superlattice-base and GaAsSb bulk-base structures," (in English), Semiconductors, vol. 44, no. 8, pp. 1096–1100, 2010/08/01 2010, doi: 10.1134/S1063782610080257.
[12] A. Dehzangi, R. McClintock, D. Wu, A. Haddadi, R. Chevallier, and M. Razeghi, "Extended short wavelength infrared het-erojunction phototransistors based on type II superlattices," Applied Physics Letters, vol. 114, no. 19, p. 191109, 2019, doi: 10.1063/1.5093560.
[13] A. Haddadi, S. Adhikary, A. Dehzangi, and M. Razeghi, "Mid-wavelength infrared heterojunction phototransistors based on type-II InAs/AlSb/GaSb superlattices," Applied Physics Letters, vol. 109, no. 2, p. 021107, 2016, doi: 10.1063/1.4958715.
[14] A. Dehzangi, R. McClintock, A. Haddadi, D. Wu, R. Chevallier, and M. Razeghi, "Type–II superlattices base visible/extended short–wavelength infrared photodetectors with a bandstructure–engineered photo–generated carrier extractor," Scientific Reports, vol. 9, no. 1, p. 5003, 2019/03/21 2019, doi: 10.1038/s41598-019-41494-6.
[15] G. G. Zegrya and A. D. Andreev, "Mechanism of suppression of Auger recombination processes in type-II heterostructures," Applied Physics Letters, vol. 67, no. 18, pp. 2681–2683, 1995, doi: doi:http://dx.doi.org/10.1063/1.114291.
[16] N. Binh-Minh, C. Guanxi, H. Minh-Anh, and M. Razeghi, "Growth and Characterization of Long-Wavelength Infrared Type-II Superlattice Photodiodes on a 3-in GaSb Wafer," Quantum Electronics, IEEE Journal of, vol. 47, no. 5, pp. 686–690, 2011, doi: 10.1109/jqe.2010.2103049.
[17] B.-M. Nguyen, D. Hoffman, P.-Y. Delaunay, E. K.-W. Huang, M. Razeghi, and J. Pellegrino, "Band edge tunability of M-structure for heterojunction design in Sb based type II superlattice photodiodes," Applied Physics Letters, vol. 93, no. 16, p. 163502, 2008, doi: doi:http://dx.doi.org/10.1063/1.3005196.
[18] P. C. Klipstein et al., "Type II Superlattice Infrared Detector Technology at SCD," Journal of Electronic Materials, vol. 47, no. 10, pp. 5725–5729, 2018/10/01 2018, doi: 10.1007/s11664-018-6527-8.
[19] A. M. Hoang, G. Chen, A. Haddadi, S. Abdollahi Pour, and M. Razeghi, "Demonstration of shortwavelength infrared pho-todiodes based on type-II InAs/GaSb/AlSb superlattices," Applied Physics Letters, vol. 100, no. 21, p. 211101, 2012, doi: doi:http://dx.doi.org/10.1063/1.4720094.
[20] R. Chevallier, A. Dehzangi, A. Haddadii, and M. Razeghi, "Type-II superlattice-based extended short-wavelength infrared focal plane array with an AlAsSb/GaSb superlattice etch-stop layer to allow near-visible light detection," Opt. Lett., vol. 42, no. 21, pp. 4299–4302, 2017/11/01 2017, doi: 10.1364/OL.42.004299.
[21] L. Höglund et al., "Influence of radiative and non-radiative recombination on the minority carrier lifetime in midwave infrared InAs/InAsSb superlattices," Applied Physics Letters, vol. 103, no. 22, p. 221908, 2013, doi: doi:http://dx.doi.org/10.1063/1.4835055.
[22] E. H. Steenbergen et al., "Significantly improved minority carrier lifetime observed in a long-wavelength infrared III-V type-II superlattice comprised of InAs/InAsSb," Applied Physics Letters, vol. 99, no. 25, p. 251110, 2011, doi: doi:http://dx.doi.org/10.1063/1.3671398.
[23] D. Z. Ting et al., "The effect of p-doping in mid- and long-wavelength InAs/InAsSb superlattice complementary barrier in-frared detectors," Applied Physics Letters, vol. 125, no. 19, 2024, doi: 10.1063/5.0236130.
[24] A. Soibel et al., "Hole diffusion length and mobility of a long wavelength infrared InAs/InAsSb type-II superlattice nBn de-sign," Applied Physics Letters, vol. 125, no. 17, 2024, doi: 10.1063/5.0236096.
[25] A. Haddadi, A. Dehzangi, R. Chevallier, S. Adhikary, and M. Razeghi, "Bias–selectable nBn dual–band long–/very long–wavelength infrared photodetectors based on InAs/InAs1−xSbx/AlAs1−xSbx type–II superlattices," Scientific Reports, vol. 7, no. 1, p. 3379, 2017/06/13 2017, doi: 10.1038/s41598-017-03238-2.
[26] A. Haddadi, A. Dehzangi, S. Adhikary, R. Chevallier, and M. Razeghi, "Background-limited long wavelength infrared InAs/InAs1-xSbx type-II superlattice-based photodetectors operating at 110K," APL Materials, vol. 5, no. 3, 2017.
[27] P. C. Klipstein, "XBnn and XBpp infrared detectors," Journal of Crystal Growth, vol. 425, pp. 351–356, 2015, doi: http://dx.doi.org/10.1016/j.jcrysgro.2015.02.075.
[28] P. C. Klipstein, "Structure of the quantum spin Hall states in HgTe/CdTe and InAs/GaSb/AlSb quantum wells," Physical Review B, vol. 91, no. 3, p. 035310, 01/20/ 2015, doi: 10.1103/PhysRevB.91.035310.
[29] T.-Y. Tsai, K. Michalczewski, P. Martyniuk, C.-H. Wu, and Y.-R. Wu, "Application of localization landscape theory and the k · p model for direct modeling of carrier transport in a type II superlattice InAs/InAsSb photoconductor system," Journal of Ap-plied Physics, vol. 127, no. 3, 2020, doi: 10.1063/1.5131470.
[30] S. Yang et al., "Enabling Ab Initio Material Design of $\mathrm{In}\mathrm{As}/\mathrm{Ga}\mathrm{Sb}$ Superlattices for Infrared Detection," Physical Review Applied, vol. 18, no. 2, p. 024058, 08/22/ 2022, doi: 10.1103/PhysRevApplied.18.024058.
[31] A. Haji Ebrahim Zargar, A. Amini, and A. Ayatollahi, "Estimation enhancing in optoelectronic property: A novel approach using orbital interaction parameters and tight-binding," Micro and Nanostructures, vol. 189, p. 207817, 2024/05/01/ 2024, doi: https://doi.org/10.1016/j.micrna.2024.207817.

[32] S. Sarkhoshan and M. A. Karami, "AlN/GaN Superlattice band structure calculation Using Empirical Tight Binding Model," Micro and Nanostructures, vol. 193, p. 207919, 2024/09/01/ 2024, doi: https://doi.org/10.1016/j.micrna.2024.207919.
[33] X. Zhu et al., "Modeling of temperature effects on band structure in type-II superlattices using an empirical tight-binding method," Applied Physics A, vol. 128, no. 7, p. 599, 2022/06/21 2022, doi: 10.1007/s00339-022-05740-5.
[34] A. Dehzangi, J. Li, and M. Razeghi, "Band-structure-engineered high-gain LWIR photodetector based on a type-II superlat-tice," Light: Science & Applications, vol. 10, no. 1, p. 17, 2021/01/14 2021, doi: 10.1038/s41377-020-00453-x.
[35] J. Smutna, R. M. Fogarty, M. R. Wenman, and A. P. Horsfield, "Systematic development of ab initio tight-binding models for hexagonal metals," Physical Review Materials, vol. 4, no. 4, p. 043801, 04/10/ 2020, doi: 10.1103/PhysRevMaterials.4.043801.
[36] Q. Gu, Z. Zhouyin, S. K. Pandey, P. Zhang, L. Zhang, and W. E, "Deep learning tight-binding approach for large-scale elec-tronic simulations at finite temperatures with ab initio accuracy," Nature Communications, vol. 15, no. 1, p. 6772, 2024/08/08 2024, doi: 10.1038/s41467-024-51006-4.
[37] L. Hao and C. S. Ting, "Minimal tight-binding model for the distinct magnetic orders of semihydrogenated and semifluori-nated graphene," Physical Review B, vol. 108, no. 24, p. 245120, 12/07/ 2023, doi: 10.1103/PhysRevB.108.245120.
[38] Y. Wei and M. Razeghi, "Modeling of type-II InAs/GaSb superlattices using an empirical tight-binding method and interface engineering," Physical Review B, vol. 69, no. 8, p. 085316, 2004. [Online]. Available: http://link.aps.org/doi/10.1103/PhysRevB.69.085316.
[39] S. Abdollahi Pour, B. Movaghar, and M. Razeghi, "Tight-binding theory for the thermal evolution of optical band gaps in semiconductors and superlattices," Physical Review B, vol. 83, no. 11, p. 115331, 03/28/ 2011, doi: 10.1103/PhysRevB.83.115331.
[40] Y. Wei, M. Razeghi, G. J. Brown, and M. Z. Tidrow, "Modeling type-II InAs/GaSb superlattices using empirical tight-binding method: new aspects," in Quantum Sensing and Nanophotonic Devices, 2004, vol. 5359: SPIE, pp. 301–308.
[41] A.-L. Phan, A. Pecchia, A. Di Vito, and M. Auf der Maur, "Empirical tight-binding method for large-supercell simulations of disordered semiconductor alloys," Physica Scripta, vol. 99, no. 7, p. 075903, 2024/05/31 2024, doi: 10.1088/1402-4896/ad4f65.
[42] A.-L. Phan, D. Soccodato, A. Pecchia, A. Di Vito, and M. Auf der Maur, "Obtaining highly transferable parameters for em-pirical tight-binding simulations: From pure materials to polytypic heterostructures and dilute alloys," Physical Review B, vol. 112, no. 16, p. 165202, 10/16/ 2025, doi: 10.1103/g9cz-c3js.
[43] A. Dehzangi, J. Li, and M. Ghasemi, "Empirical tight binding method for quantum mechanical approach of band structure computation," in Quantum Computing, Communication, and Simulation VI, 2026, vol. 13919: SPIE, pp. 152–160.
[44] A. Dehzangi and J. Li, "Empirical Tight Binding Modeling of InAs/InAs1-xSbx Ternary-Based Strain-Balanced Superlattices," in 2025 IEEE 25th International Conference on Nanotechnology (NANO), 13–16 July 2025 2025, pp. 156–161, doi: 10.1109/NANO63165.2025.11113673.
[45] A. Dehzangi and J. Li, "Modeling of InAs/InAs1-xSbx strained-layer superlattice using a modified empirical tight binding method," IEEE Journal of Selected Topics in Quantum Electronics, pp. 1–10, 2025, doi: 10.1109/JSTQE.2025.3590792.
[46] J. C. Slater and G. F. Koster, "Simplified LCAO Method for the Periodic Potential Problem," Physical Review, vol. 94, no. 6, pp. 1498–1524, 06/15/ 1954. [Online]. Available: http://link.aps.org/doi/10.1103/PhysRev.94.1498.
[47] P. O. Löwdin, "On the Non-Orthogonality Problem Connected with the Use of Atomic Wave Functions in the Theory of Molecules and Crystals," The Journal of Chemical Physics, vol. 18, no. 3, pp. 365–375, 1950, doi: 10.1063/1.1747632.
[48] P. Vogl, H. P. Hjalmarson, and J. D. Dow, "A Semi-empirical tight-binding theory of the electronic structure of semicon-ductors†," Journal of Physics and Chemistry of Solids, vol. 44, no. 5, pp. 365–378, 1983/01/01/ 1983, doi: https://doi.org/10.1016/0022-3697(83)90064-1.
[49] W. A. Harrison, "Electronic Structure and the Properties of Solids - The Physics of the Chemical Bond," ed: Dover Publica-tions, 1989.
[50] A. Khoshakhlagh, E. Plis, S. Myers, Y. D. Sharma, L. R. Dawson, and S. Krishna, "Optimization of InAs/GaSb type-II super-lattice interfaces for long-wave (~8μm) infrared detection," Journal of Crystal Growth, vol. 311, no. 7, pp. 1901–1904, 2009/03/15/ 2009, doi: https://doi.org/10.1016/j.jcrysgro.2008.11.027.
[51] A. M. Hoang, "Theoretical design and material growth of Type-II Antimonide-based superlattices for multi-spectral infrared detection and imaging," Ph.D., Northwestern University, United States -- Illinois, 10160447, 2016.
[52]D. Wu, A. Dehzangi, and M. Razeghi, "Demonstration of mid-wavelength infrared nBn photodetectors based on type-II InAs/InAs1-xSbx superlattice grown by metal-organic chemical vapor deposition," Applied Physics Letters, vol. 115, no. 6, p. 061102, 2019, doi: 10.1063/1.5100617.
[53] N. J. Ekins-Daukes et al., "Strain-balanced quantum well solar cells," Physica E: Low-dimensional Systems and Nanostruc-tures, vol. 14, no. 1, pp. 132–135, 2002/04/01/ 2002, doi: https://doi.org/10.1016/S1386-9477(02)00378-8.
[54] N. J. Ekins-Daukes, K. Kawaguchi, and J. Zhang, "Strain-Balanced Criteria for Multiple Quantum Well Structures and Its Signature in X-ray Rocking Curves," Crystal Growth & Design, vol. 2, no. 4, pp. 287–292, 2002, doi: 10.1021/cg025502y.
[55] P. F. Fewster, X-Ray scattering from semiconductors and other materials. World Scientific, 2015.
[56] F. Rahayu, "HRXRD characterization of MBE epitaxial materials and related nanostructures," 2015.
[57] N. J. Ekins-Daukes, K. Kawaguchi, and J. Zhang, "Strain-Balanced Criteria for Multiple Quantum Well Structures and Its Signature in X-ray Rocking Curves†," Crystal Growth & Design, vol. 2, no. 4, pp. 287–292, 2002/07/01 2002, doi: 10.1021/cg025502y.
[58] L. Vegard, "Die Konstitution der Mischkristalle und die Raumfüllung der Atome," (in German), Z. Physik, vol. 5, no. 1, pp. 17–26, 1921/01/01 1921, doi: 10.1007/BF01349680.
[59] G. Klimeck, R. Chris Bowen, T. B. Boykin, and T. A. Cwik, "Tight-binding parameters for transport simulations in compound semiconductors," Superlattices and Microstructures, vol. 27, no. 5–6, pp. 519–524, 5// 2000, doi: http://dx.doi.org/10.1006/spmi.2000.0862.
[60] W. A. Harrison and S. Ciraci, "Bond-orbital model. II," Physical Review B, vol. 10, no. 4, pp. 1516–1527, 1974, doi: 10.1103/PhysRevB.10.1516.
[61] S. T. Pantelides and W. A. Harrison, "Structure of the valence bands of zinc-blende-type semiconductors," Physical Review B, vol. 11, no. 8, pp. 3006–3021, 04/15/ 1975. [Online]. Available: http://link.aps.org/doi/10.1103/PhysRevB.11.3006.
[62] S. Froyen and W. A. Harrison, "Elementary prediction of linear combination of atomic orbitals matrix elements," Physical Review B, vol. 20, no. 6, pp. 2420–2422, 09/15/ 1979. [Online]. Available: http://link.aps.org/doi/10.1103/PhysRevB.20.2420.
[63] Y. P. Varshni, "Temperature dependence of the energy gap in semiconductors," Physica, vol. 34, no. 1, pp. 149–154, // 1967, doi: http://dx.doi.org/10.1016/0031-8914(67)90062-6.

[64] M. Cardona and N. E. Christensen, "Acoustic deformation potentials and heterostructure band offsets in semiconductors," Physical Review B, vol. 35, no. 12, pp. 6182–6194, 04/15/ 1987. [Online]. Available: http://link.aps.org/doi/10.1103/PhysRevB.35.6182.

[65] R. L. Anderson, "Experiments on Ge-GaAs heterojunctions," Solid-State Electronics, vol. 5, no. 5, pp. 341–351, 9// 1962, doi: http://dx.doi.org/10.1016/0038-1101(62)90115-6.

[66] C. G. Van de Walle, "Band lineups and deformation potentials in the model-solid theory," Physical Review B, vol. 39, no. 3, pp. 1871–1883, 01/15/ 1989. [Online]. Available: http://link.aps.org/doi/10.1103/PhysRevB.39.1871.

[67] A. Dehzangi and J. Li, "Quantum Mechanical Approach for Modeling of Ternary Based Strained-Layer Superlattice," arXiv preprint arXiv:2507.12813, 2025.

[68] J. A. Van Vechten and T. K. Bergstresser, "Electronic Structures of Semiconductor Alloys," Physical Review B, vol. 1, no. 8, pp. 3351–3358, 04/15/ 1970. [Online]. Available: http://link.aps.org/doi/10.1103/PhysRevB.1.3351.

[69] A. B. Chen and A. Sher, "Electronic structure of pseudobinary semiconductor alloys A Al$x$ Ga1−$x$⊠As, Ga⊠P$x$⊠As1−$x$, and Ga$x$⊠In1−$x$⊠P," Physical Review B, vol. 23, no. 10, pp. 5360–5374, 05/15/ 1981.

[70] N. Tit, I. M. Obaidat, and H. Alawadhi, "Origins of bandgap bowing in compound-semiconductor common-cation ternary alloys," Journal of Physics: Condensed Matter, vol. 21, no. 7, p. 075802, 2009. [Online]. Available: http://stacks.iop.org/0953-8984/21/i=7/a=075802.

[71] D. Olguin, R. Baquero, and R. de Coss, "The band gap of II-VI ternary alloys in a tight-binding description " Revista Mexicana de Fisica, vol. 47, no. 1, pp. 43–49, 2001.

[72] I. Vurgaftman, J. R. Meyer, and L. R. Ram-Mohan, "Band parameters for III–V compound semiconductors and their alloys," Journal of Applied Physics, vol. 89, no. 11, pp. 5815–5875, 2001, doi: doi:http://dx.doi.org/10.1063/1.1368156.

[73] J. M. Moison, C. Guille, F. Houzay, F. Barthe, and M. Van Rompay, "Surface segregation of third-column atoms in group III-V arsenide compounds: Ternary alloys and heterostructures," Physical Review B, vol. 40, no. 9, pp. 6149–6162, 09/15/ 1989. [Online]. Available: http://link.aps.org/doi/10.1103/PhysRevB.40.6149.

[74] K. Muraki, S. Fukatsu, Y. Shiraki, and R. Ito, "Surface segregation of In atoms during molecular beam epitaxy and its influence on the energy levels in InGaAs/GaAs quantum wells," Applied Physics Letters, vol. 61, no. 5, pp. 557–559, 1992, doi: doi:http://dx.doi.org/10.1063/1.107835.

[75] J. F. Zheng, J. D. Walker, M. B. Salmeron, and E. R. Weber, "Interface segregation and clustering in strained-layer In-GaAs/GaAs heterostructures studied by cross-sectional scanning tunneling microscopy," Physical Review Letters, vol. 72, no. 15, pp. 2414–2417, 04/11/ 1994. [Online]. Available: http://link.aps.org/doi/10.1103/PhysRevLett.72.2414.

[76] E. Bellotti, F. Bertazzi, A. Tibaldi, J. Schuster, J. Bajaj, and M. Reed, "Disorder-Induced Degradation of Vertical Carrier Transport in Strain-Balanced Antimony-Based Superlattices," Physical Review Applied, vol. 16, no. 5, p. 054028, 11/15/ 2021, doi: 10.1103/PhysRevApplied.16.054028.

[77] H. J. Haugan, G. J. Brown, and J. A. Peoples, "On the study of antimony incorporation in InAs/InAsSb superlattices for in-frared sensing," Journal of Vacuum Science & Technology B, vol. 35, no. 2, 2017, doi: 10.1116/1.4977009.

[78] J. Steinshnider, J. Harper, M. Weimer, C. H. Lin, S. S. Pei, and D. H. Chow, "Origin of Antimony Segregation in $\mathrm{GaInSb}/\mathrm{InAs}$ Strained-Layer Superlattices," Physical Review Letters, vol. 85, no. 21, pp. 4562–4565, 11/20/ 2000. [Online]. Available: http://link.aps.org/doi/10.1103/PhysRevLett.85.4562.

[79] M. S. Milosavljevic, P. T. Webster, and S. R. Johnson, "Impact of unintentional Sb in the tensile InAs layer of strain-balanced type-II InAs/InAsSb superlattices grown on GaSb by molecular beam epitaxy," Journal of Applied Physics, vol. 134, no. 5, 2023, doi: 10.1063/5.0161046.

[80] Q. Yang, R. Yuan, L. Wang, R. Shi, and J.-M. Zuo, "Antimony segregation in an InAs/InAs1−xSbx superlattice grown by metalorganic chemical vapor deposition," Journal of Applied Physics, vol. 130, no. 9, 2021, doi: 10.1063/5.0060777.

[81] A. M. Hoang, G. Chen, R. Chevallier, A. Haddadi, and M. Razeghi, "High performance photodiodes based on InAs/InAsSb type-II superlattices for very long wavelength infrared detection," Applied Physics Letters, vol. 104, no. 25, p. 251105, 2014, doi: doi:http://dx.doi.org/10.1063/1.4884947.

[82] A. Haddadi, R. Chevallier, G. Chen, A. Hoang, and M. Razeghi, "Bias-selectable dual-band mid-/long-wavelength infrared photodetectors based on InAs/InAs1− xSbx type-II superlattices," Applied Physics Letters, vol. 106, no. 1, p. 011104, 2015.

[83] M. Razeghi, A. Haddadi, A. Dehzangi, R. Chevallier, and T. Yang, "Recent advances in InAs/InAs1-xSbx/AlAs1-xSbx gap-engineered type-II superlattice-based photodetectors," in Proc. SPIE 10177, Infrared Technology and Applications XLIII, 1017705 2017, vol. 10177: SPIE, pp. 1017705–1017705–11. [Online]. Available: http://dx.doi.org/10.1117/12.2267044. [Online]. Available: http://dx.doi.org/10.1117/12.2267044

[84] A. Haddadi, G. Chen, R. Chevallier, A. M. Hoang, and M. Razeghi, "InAs/InAs1−xSbx type-II superlattices for high perfor-mance long wavelength infrared detection," Applied Physics Letters, vol. 105, no. 12, 2014, doi: 10.1063/1.4896271.

[85] M. Razeghi, A. Haddadi, A. M. Hoang, R. Chevallier, S. Adhikary, and A. Dehzangi, "InAs/InAs1-xSbx type-II superlattices for high performance long wavelength infrared detection," in Proc. SPIE 9819, Infrared Technology and Applications XLII, 2016, vol. 9819, pp. 981909–981909–8, doi: 10.1117/12.2228306. [Online]. Available: http://dx.doi.org/10.1117/12.2228306

[86] F. Hatami and O. Bierwagen, "3.13 - Growth of Low-Dimensional Semiconductors Structures," in Comprehensive Semi-conductor Science and Technology, P. Bhattacharya, R. Fornari, and H. Kamimura Eds. Amsterdam: Elsevier, 2011, pp. 523–583.

[87] A. Haddadi et al., "High-performance short-wavelength infrared photodetectors based on type-II InAs/InAs1-xSbx/AlAs1−xSbx superlattices," Applied Physics Letters, vol. 107, no. 14, 2015, doi: 10.1063/1.4932518.

[88] M. Razeghi et al., "High-performance short-wavelength infrared photodetectors based on type-II InAs/InAs1-xSbx/AlAs1-xSbx superlattices," in Proc. SPIE 9819, Infrared Technology and Applications XLII, 2016, vol. 9819: SPIE, pp. 98190A–98190A–7. [Online]. Available: http://dx.doi.org/10.1117/12.2228611. [Online]. Available: http://dx.doi.org/10.1117/12.2228611